\documentclass[conference]{IEEEtran}
\ifCLASSINFOpdf
\else
\fi
\usepackage{amsthm}
\usepackage{makeidx}
\usepackage{examples}
\usepackage[pdftex]{graphicx}
\usepackage{amssymb}
\usepackage{graphicx}
\usepackage{algorithm}
\usepackage{epstopdf}
\usepackage{color}
\usepackage{url}
\graphicspath{{../eps/}{../ps/}}
\usepackage{psfrag}
\usepackage{algorithm}
\usepackage{algorithmic}
\usepackage{amsmath}
\usepackage{hyperref}
\newtheorem{theorem}{Theorem}

\begin{document}
%
% paper title
% can use linebreaks \\ within to get better formatting as desired
\title{RNA as a Permutation}

% author names and affiliations
% use a multiple column layout for up to three different
% affiliations
\author{
\IEEEauthorblockN{Nilay Chheda}
\IEEEauthorblockA{Qualcomm India\\
 Hyderabad, India\\
Email:nchheda@qti.qualcomm.com}
\and
\IEEEauthorblockN{Manish K. Gupta}
\IEEEauthorblockA{Laboratory of Natural Information Processing\\
Dhirubhai Ambani Institute of Information\\ and Communication Technology\\
Gandhinagar, Gujarat, 382007 India\\
Email: mankg@computer.org}
}

% conference papers do not typically use \thanks and this command
% is locked out in conference mode. If really needed, such as for
% the acknowledgment of grants, issue a \IEEEoverridecommandlockouts
% after \documentclass

% for over three affiliations, or if they all won't fit within the width
% of the page, use this alternative format:
%
%\author{\IEEEauthorblockN{Michael Shell\IEEEauthorrefmark{1},
%Homer Simpson\IEEEauthorrefmark{2},
%James Kirk\IEEEauthorrefmark{3},
%Montgomery Scott\IEEEauthorrefmark{3} and
%Eldon Tyrell\IEEEauthorrefmark{4}}
%\IEEEauthorblockA{\IEEEauthorrefmark{1}School of Electrical and Computer Engineering\\
%Georgia Institute of Technology,
%Atlanta, Georgia 30332--0250\\ Email: see http://www.michaelshell.org/contact.html}
%\IEEEauthorblockA{\IEEEauthorrefmark{2}Twentieth Century Fox, Springfield, USA\\
%Email: homer@thesimpsons.com}
%\IEEEauthorblockA{\IEEEauthorrefmark{3}Starfleet Academy, San Francisco, California 96678-2391\\
%Telephone: (800) 555--1212, Fax: (888) 555--1212}
%\IEEEauthorblockA{\IEEEauthorrefmark{4}Tyrell Inc., 123 Replicant Street, Los Angeles, California 90210--4321}}

% use for special paper notices
%\IEEEspecialpapernotice{(Invited Paper)}

% make the title area
\maketitle

\begin{abstract}
%\boldmath
RNA secondary structure prediction and classification are two important problems in the field of RNA biology. Here, we propose a new permutation based approach to create logical non-disjoint clusters of different secondary structures of a single class or type. Many different types of techniques exist to classify RNA secondary structure data but none of them have ever used permutation based approach which is very simple and yet powerful. We have written a small JAVA program to generate permutation, apply our algorithm on those permutations and analyze the data and create different logical clusters. We believe that these clusters can be utilized to untangle the mystery of RNA secondary structure and analyze the development patterns of unknown RNA. 
\end{abstract}
% IEEEtran.cls defaults to using nonbold math in the Abstract.
% This preserves the distinction between vectors and scalars. However,
% if the conference you are submitting to favors bold math in the abstract,
% then you can use LaTeX's standard command \boldmath at the very start
% of the abstract to achieve this. Many IEEE journals/conferences frown on
% math in the abstract anyway.
% no keywords
% For peer review papers, you can put extra information on the cover
% page as needed:
% \ifCLASSOPTIONpeerreview
% \begin{center} \bfseries EDICS Category: 3-BBND \end{center}
% \fi
%
% For peerreview papers, this IEEEtran command inserts a page break and
% creates the second title. It will be ignored for other modes.
\IEEEpeerreviewmaketitle

%%%%%%%%%%%%%%%%%%%%%%%%%%%%
\section{Introduction}
RNA molecule is one of the three major macro molecules which is essential for all existing models of human life. RNA is a short form of Ribo nucleic acid. RNA molecules plays pivotal role in many biological functions. RNA can rebuild and transport genetic data~\cite{citeulike:137111}, drive chemical reactions~\cite{lilley2005structure} and administer gene expressions~\cite{citeulike:486661}. RNA molecule's capability to perform bio-molecular computation through nanotechnology has escalated its importance among researchers from various fields (~\cite{citeulike:2645587},~\cite{citeulike:7983182} and~\cite{sullenger2000series} ). 
RNAs form primary, secondary and tertiary structures very much like DNA and protein. In essence, primary structure is a simple one dimensional sequence of nucleotides whereas secondary and tertiary structures are nothing but two dimensional and three dimensional representation of that sequence respectively. There has been a debate going on for many years whether RNA is the only molecule that is responsible for evolution of life or RNA along with DNA and protein have facilitated the evolution. 

%%%%%
\subsection{Concept of RNA World}
"RNA world" is theoretical time of the early ecosphere. During this time period, knowledge required for life and alchemical movement of lively organisms were accommodated by RNA molecules~\cite{citeulike:2085297}. Another set of arguments indicated that compared to RNA, availability of DNA was rich in that time period. DNA is more stable in mildly primitive atmosphere. Existence of ammonia-rich sea and the seasoning of primitive crusts of earth also support former statement~\cite{citeulike:8506292}. Another variant postulated that a different kind of nucleic acid came first which was called pre-RNA. It had a property to emulate itself and RNA, as we know it today, substituted it eventually. On contrary, new results prompted that pyrimidine ribonucleotides can be prepared under certain prebiotic setup~\cite{citeulike:4510662}.

%%%%%
\subsection{Types of RNA}
Majority variants of RNA fall under RNAs that are either involved in protein synthesis or DNA replication or Regulatory RNA. Some RNA has the double stranded structure like DNA they are named as dsRNA. Non-coding RNA (ncRNA) is one which does not translate itself into protein. Often it is supposed that large amount of genetic information is carried out by proteins. Modern studies have implied that unlike proteins, ncRNAs are generated by the translation of genome of many mammals and certain organisms~\cite{citeulike:612596}. Many regulatory RNA, tRNA, rRNA also fall under ncRNA category. Three major RNA involved in protein synthesis are following:
\begin{enumerate}
\item \textbf{mRNA} -- mRNA stands for Messenger RNA. mRNA plays role to forward genetic information from DNA to the ribosome. In messages transferred, mRNA encodes details about amino acid strand for gene expression of protein derivatives. It is a one of the largest subset of RNA.
\item \textbf{rRNA} -- rRNA stands for Ribosomal ribo nucleic acid. It is one of the important constituent of ribosome and fundamental for protein construction in all forms of life.
\item \textbf{tRNA} -- tRNA stands for Transfer RNA. As the name suggests, it acts as an interconnection between the different types of chain of nucleotides (DNA \& RNA) and amino acid string of proteins.
\end{enumerate}

We do not know the correct answer of RNA world debate till date but what we know is that "RNA", irrespective of the fact whether it is the only molecule responsible for evolution of life, is certainly responsible for some of the biological functions. This very fact is motivating enough to analyze its behavior. Its behavior can be understood by finding patterns in its secondary structure in which it folds. This folding can be cumulative result of many different known and unknown biological, chemical, thermodynamic and mathematical parameters. Here we have tried to explore mathematical aspect of RNA secondary structures in the form of permutation.

After giving basic introduction in section $I$, we explain basics about RNA structure and different techniques to represent those structures in section $II$ and $III$ respectively. We also introduce our own representation of RNA secondary structure which is derived from one of the existing representation in section $III$. In section $IV$, we  explain our algorithm to calculate two different types of similarity score based on the new representation that we proposed in the section $III$. We also show results of our analysis based on the algorithm. In section $V$, we discuss what more can be build up based on the results that we got. Finally, we conclude our work in the section $VI$.

%%%%%%%%%%%%%%%%%%%%%%%%%
\section{RNA Structure}
As we mentioned earlier, RNA has three different types of structure. One dimensional structure is called primary whereas two and three dimensional structures are called secondary and tertiary respectively. Primary one is a random linear sequence of four nucleotide namely -- Adenine, Cytosine, Guanine and Uracil (represented as alphabets A, C, G and U respectively). RNA is usually single stranded unlike DNA.
%%%%%%%%%%%%%%%%%%%
\subsection{Chemical Structure}
Chemically, RNA is a linear chain of polymers of ribose sugar having ringed structure. This ring has five carbons. Third ($3'$) and Fifth ($5'$) carbon are connected with phosphate group which act as a linkage forming chain of ribose sugar. First carbon is connected with one of the four base group (A,C,G \& U) which are derivatives of purines and pyrimidines chemical structure. Now, this base group can pair with another base group with hydrogen bonds. A \& T can be paired with two hydrogen bonds and C \& U can be paired with three hydrogen bonds. Double stranded RNA (dsRNA) and DNA have complementary linear sequence means sequence which can be paired with original sequence with Watson-Crick pairing. Usually original sequence is observed from ($3'$) end to ($5'$) end and vice verse for complementary sequence. These relationships are called the rules of Watson-Crick base pairing~\cite{citeulike:5432573}. There is a thermodynamic energy involved with each hydrogen bond according to which relationships between A \& T and C \& G are the most stable. Sometimes G is also paired with U with two hydrogen bond. This relationship is called as wobble pairing. This type of pairing is comparatively much less in numbers than Watson-Crick pairing.

%%%%%%%%%%%%%%%%%%%%%%%%%
\subsection{Secondary \& Tertiary Structure}
\begin{figure}[here]
\includegraphics{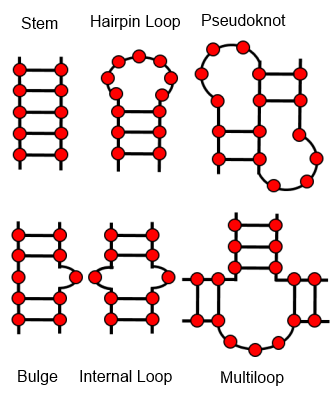}
\caption{Different types of RNA Secondary Structures. [1] First structure represents a stem with five paired bases. [2] Second structure represents Hairpin loop having three paired and six unpaired bases. [3] Third structure is of a HH type of pseudoknot which is one of the most common type in which unpaired base of two hairpins structure get paired with the other one. [4] In the first structure in the bottom row, we have shown a bulge of size one. There can be multiple unpaired bases in structures like this. [5] In fifth structrue we have shown internal loop which usually have double bulge like a structure. [6] In last image in the bottom right we have shown multiloop which joins multiple substructure.}
\label{fig:secondaryStructure}
\end{figure}

As discussed in the previous subsection, RNA strand consist of bases which can pair with each other. This pairing take place in a way that the resulting structure becomes the most stable in thermodynamic aspects. There are only few possible substructure shapes in which RNA can fold. There can be multiple substructures of same type in single secondary structure and if not same they can be different in size (in terms of no. of paired and unpaired base). As shown in Figure~\ref{fig:secondaryStructure}, major substructure types are Stem, Bulge, Hairpin,  Multiloop (Junction), Interior Loop and Pseudoknot. Any RNA secondary structure will be combination of these substructures only.

Secondary structure of an RNA is nothing but a planar view of actual real three dimensional (tertiary) structure. This tertiary structure is the actual mystery that many researchers want to solve but only possible way to understand it is by understanding two dimensional structure of it first. As shown in Figure~\ref{fig:secondaryStructure}, psuedoknots are the most complex and hardest to predict substructures because of its bonding nature. Psuedoknots are formed when two substructure other than pseudoknot come closer in three dimensional geometry and certain portion of strand in one substructure is complementary to some portion of strand in the other structure. Usually, unpaired portion of both the substructures are paired when psuedoknots are formed.
\subsection{Secondary Structure Prediction}
The Human Genome Project which started in 1993 by the global research community have finished sequencing approximately 200 million base pairs by end of its first five year plan which ended in 1998~\cite{citeulike:7846921}. This kind of large scale project along with many similar small scale projects have generated immense amount of biological data in past few years. Now, newer challenge standing ahead of research community is to understand this information. In recent years importance of understanding secondary structure of RNA has become very rewarding and essential work. Efforts to accomplish this task has resulted in various techniques and approaches. RNA secondary structures are critical in many biological mechanisms and accurate techniques for prediction can provide crucial pathway to conduct effective research in the area~\cite{citeulike:7273246}.

First breakthrough algorithm was introduced in 1978 which was based on dynamic programming paradigm and designed to do sequence matching~\cite{citeulike:2599605}. Based on this algorithm, efficient and precise method to predict secondary structure was designed in 1980~\cite{citeulike:547969}.This algorithm is famously known as "Nussinov's algorithm". Now problem with this method was that it did not take into account thermodynamic and energy related factors affecting structure. Micheal Zucker and his colleagues came up with newer algorithm which was able to predict the energetically most stable structures of an RNA~\cite{citeulike:6760869}. They also implemented this algorithm as a computer software~\cite{citeulike:599795}. They have been constantly improving their algorithm and software based on better free energy calculations~\cite{citeulike:118766}. As new standards and formats for representing RNA secondary structure were developed e.g. RNAML~\cite{citeulike:400898}, they also updated their software to provide output in this standard format. As the usage of internet grew in early 20th century, they set up web server which provided their software as an online service. They called this server as mFold server~\cite{citeulike:100029}. This server, till date, is the most popular server for structure prediction due to its wide range of features and accuracy. Meanwhile another similar effort was carried out by group of researchers who also tried to develop a computationally faster techniques for prediction and comparison of different RNA structures~\cite{citeulike:339293}. They also set up a web server at almost the same time as mFold server~\cite{citeulike:118727}. Their server is called a Vienna Web Server.

Major shortcoming of Zucker's algorithm was that it was not able to predict pseudoknot within a secondary structure. This limitation motivated researchers to design algorithm which can also predict structures with pseudoknot as RNA pseudoknots are quiet influential for various existing RNAs~\cite{citeulike:9566547}. The very first such algorithms was developed in 1999 and named as "PKNOTS"~\cite{citeulike:118804}. This algorithm was capable of computing secondary structure containing widest class of pseudoknots but it was computationally very inefficient  ($O(n^6)$) although it was developed based on the Zucker's original algorithm which runs in $O(n^3)$ . There are different types of pseudoknots that occur in secondary structures (~\cite{citeulike:118804}, ~\cite{citeulike:118766}). There were many variants of~\cite{citeulike:118804} algorithm (~\cite{citeulike:2599621}, ~\cite{citeulike:118685}, ~\cite{citeulike:4785118}, ~\cite{citeulike:652740}) developed during the time frame of five years . Most of them discarded one or more types of pseudoknots from their methods to bring down the computational complexity. But after all these efforts computational complexity was ranging from $O(n^4)$ to $O(n^6)$ which was still better than Zucker's $O(n^3)$. Another problem with these methods was is that they were producing only the optimal solution while ignoring sub optimal solution which may reveal the true structure~\cite{citeulike:5892260}. On the contrary, computationally efficient (within $O(n^3$)) methods (~\cite{citeulike:118810}, ~\cite{citeulike:494445}) have also been developed which takes heuristic approach without restricting to any particular class of pseudoknots. These methods do not guarantee optimality and quality for the results of prediction though their computational complexity is much better than previous methods~\cite{citeulike:5892260}.

%%%%%%%%%%%%%%%%%%%%%%%%%%%%%%%%%%
\section{Representation of RNA secondary structures}
After development of so many RNA secondary structure prediction techniques, there were three main areas which emerged as a logical extension to mankind's aspiration to solve the mystery of life which are following:
\begin{enumerate}
\item \textbf{Classification of predicted structures} -- Classification brings RNA secondary structure with similar characteristic in one or more ways together. This field encouraged creation of database having basic database functions like search, sort, add, delete. This database operations can be helpful to carrying out further scientific research on the data. Database containing one dimensional sequence data of various kind of proteins, viruses, RNAs, DNAs etc. already exist. Very recent work~\cite{citeulike:10131681} has brought together almost all the one dimensional sequential data that is known to the world which is just 10\% of total possible data.
\item \textbf{File Format} -- Today, we are living in the computer age. Practically every research area in the world need computers for carrying out various tasks of computation, simulation etc. Biological data size is one of the largest in the world and large data size are nearly impossible for humans to process. RNA secondary structure data is also quiet big which has created requirement of proper file formats that computers can interpret.
\item \textbf{Display Tools} -- Computers can do computation with the data but human touch (in technical terms input) is always required to do meaningful computation. Computer needs to display data in a way that a human can understand e.g. showing secondary structure drawing along with a text file having thousands of alphanumeric characters is much more convenient \& informative for human than simply showing the same text file without visual output. Display tools developers are constantly trying to make visuals clearer and more informative.
\end{enumerate}

Some of the RNA secondary database existing today are NDB~\cite{citeulike:4936425}, Rfam~\cite{citeulike:6846542}, RNasePDB~\cite{citeulike:8318035}, sprinzIDB~\cite{citeulike:8318243}, tmRDB \&  SRPDB~\cite{citeulike:5107270}, CRW~\cite{citeulike:118653} and PDB~\cite{citeulike:1485546}. Pseudobase is a database which only contain structures having pseudoknot (~\cite{citeulike:2599486}, ~\cite{citeulike:8436659}). Based on various minimum free energy (mfe) based prediction algorithm, ~\cite{citeulike:678833} has also tried to classify pseudoknotted structures. All of these databases are constantly being updated with more data, better accessibility and useful features. Very recent example can be Rfam~\cite{citeulike:6846542} whose 11th version has just been released~\cite{citeulike:11636259}. We have used RNA STRAND~\cite{citeulike:3124365} database for our analysis purpose, as it has collaborated almost all of above mentioned databases and combined all of them together. 

Different notation used to represent secondary structures are String Notation, Bracket Dot Notation, Linked Graph Notation, Circular Notation, Dot Plot notation, Mountain Plot Notation, Mountain Metric notation, Tree Notation. Bracket notation has become quite popular and an extended dot bracket notation~\cite{ramlan2008extended} has been developed which can also represent pseudoknots. Dot bracket notation is just a sequence of dots and brackets in which dot represents unpaired base and brackets (parentheses, square brackets, and curly braces - depending upon the base pairing and structures like pseudoknot)  represent paired base. Originally dot bracket notation was used by Vienna Web Server~\cite{citeulike:118727}. RNADraw~\cite{matzura1996rnadraw} and RNAviz~\cite{citeulike:2645822} are software that can visualize secondary structures. Most of these kind of software use XML based RNAML format~\cite{citeulike:400898}. Currently, Psedoviewer~\cite{citeulike:721508} is the most popular visualization software as it can draw pseudoknotted structures. It has constantly updated its software to support more and more pseudoknots and better and efficient 2D and 3D visualization~\cite{citeulike:4544619}. One of the popular file format is ".ct". This format is nothing but the table containing six columns and along with one common header text. Common columns of the tables are index of base in the sequence, index of paired base in the sequence for corresponding base, alphabet representing base etc. There are many variants of .ct format like "Vienna .ct", "RnaViz ct", "Mac ct" and ".bpseq". All of them have table with minimum three columns which we just mentioned. Some of them have header containing energy related information and some of them have columns that contain energy related information. ".bpseq" is the file format which has just three column table and a simple header containing information like name of the file, source of the file and accession number. All of the formats are supported by mFold~\cite{citeulike:100029} and Vienna Server~\cite{citeulike:118727}. We have used ".bpseq" file format to create our hybrid notation which we use for our analysis.
%%%%%%%%%%%%%%%%%%%%%%%%%
\subsection{New Hybrid Representation}
Our new hybrid notation is developed based on the permutation concept of mathematics. In the subsection below, we will explain what permutation is,  how we are using permutation based representation for creating clusters from the given data set and finally how permutation notation can be helpful to propose a new kind of classification.
%%%%%%%%%%%%%%%%%%%%%%%
\subsection{Permutation}
Before, we explain about our hybrid notation, we would like to first explain what permutation is. Dictionary meaning of permutation is rearrangement. For $n$ number of different objects, there are $n!$ different arrangement or permutation possible. In combinations theory, a permutation is a sequence having all the elements of a finite set just once.

\begin{figure}[here]
\includegraphics{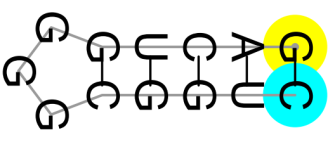}
\caption{Secondary Structure for PDB\_00226}
\label{fig:PDB00226}
\end{figure}

\begin{table}[h]
\scriptsize
\caption{Permutation based hybrid notation of RNA secondary structure}
\label{tab:newNotation}
\begin{tabular}{c|c|c|c|c|c|c|c|c|c|c|c|c}
1&2&3&4&5&6&7&8&9&10&11&12&13\\
G&A&C&U&G&G&G&G&C&G&G&U&C\\
\smallskip
13&12&11&10&9&0&0&0&5&4&3&2&1\\
\smallskip
13&12&11&10&9&6&7&8&5&4&3&2&1\\
0&1&0&0&0&0&0&0&0&0&0&0&0\\
0&0&1&0&0&0&0&0&1&0&0&0&1\\
1&0&0&0&1&1&1&1&0&1&1&0&0\\
0&0&0&1&0&0&0&0&0&0&0&1&0\\
\end{tabular}
\end{table}

In Table~\ref{tab:newNotation}, first three rows are simply horizontal representation of ".bpseq" file format of secondary structure of tRNA with accession number PDB\_00226 (shown in Figure~\ref{fig:PDB00226}) taken from~\cite{citeulike:3124365}. Fourth row is nothing but the permutation that we have derived from the third row by simply replacing $0$ with its corresponding index. Here indexes $6$, $7$ and $8$ are unpaired which is why their value in third row is $0$ unlike in row 4. This way, we can get unique permutation defined on a set $S$. Here length of RNA sequence is $13$. In fourth row integers from $1$ to $13$ appear exactly once which is why it is a permutation. Next four rows are indicator sequence~\cite{citeulike:1195219} of A, C, G and U respectively. Let us call this eight sequence notation/format/representation as "permutSeq".

Permutation defined on the RNA sequence has a very special property that permutation always form two cycles or one cycle. This is because of the nature of RNA, where there can be either paired base or unpaired base. For unpaired base we always get one cycle and for paired base we always get two cycles. 
%Let us take an example of a synthetic RNA with accession number PDB\_00590 of length 10 which forms a hairpin structure~\cite{citeulike:3124365}.

\begin{figure}[h!]
{ \scriptsize
\[ \left(
\begin{array}{ccccccccccccc}
1&2&3&4&5&6&7&8&9&10&11&12&13\\
13&12&11&10&9&6&7&8&5&4&3&2&1
\end{array}
\right) \]
}
$$ =  (1\;13)\,(2\;12)\,(3\;11)\,(4\;10)\,(5\;9)\,(6)\,(7)\,(8)$$
\caption{A permutation representation of  PDB\_00226 of Figure \ref{fig:PDB00226}}
\label{default}
\end{figure}

%%%%
\begin{figure}[h!]
\[
\begin{array}{ccc}
	\bigskip
	% First Row
	\left(
	\begin{array}{cccc}
	1&2&3&4\\
	1&2&4&3\\
	\end{array}
	\right) &
	\left(
	\begin{array}{cccc}
	1&2&3&4\\
	1&4&3&2\\
	\end{array}
	\right) &
	\left(
	\begin{array}{cccc}
	1&2&3&4\\
	1&3&2&4\\
	\end{array}
	\right)\\

	\bigskip
	% Second Row

	\left(
	\begin{array}{cccc}
	1&2&3&4\\
	4&2&3&1\\
	\end{array}
	\right) &
	\left(
	\begin{array}{cccc}
	1&2&3&4\\
	3&2&1&4\\
	\end{array}
	\right) &
	\left(
	\begin{array}{cccc}
	1&2&3&4\\
	2&1&3&4\\
	\end{array}
	\right)\\

	% Third Row

	\left(
	\begin{array}{cccc}
	1&2&3&4\\
	2&1&4&3\\
	\end{array}
	\right) &
	\left(
	\begin{array}{cccc}
	1&2&3&4\\
	3&4&1&2\\
	\end{array}
	\right) &
	\left(
	\begin{array}{cccc}
	1&2&3&4\\
	4&3&2&1\\
	\end{array}
	\right)\\
\end{array}
\]
\caption{All 9 possible permutations of length 4 used in counting}
\label{fig:cases}
\end{figure}
One can count such permutations. Suppose, there is an RNA sequence of length $3$. If we hypothetically enumerate all the possible permutation of length 3, there will be some paired bases and some unpaired bases. Paired bases will always be in even number as they need a partner to pair with. So, essentially it is a partition problem. $3$ can be partitioned into unpaired and paired ($u$, $p$) as $(0,3)$ or $(1,2)$. For $(0,3)$ there will be only one possible structure. For $(1,2)$, there will be three possible structures calculated as ${3\choose1} \times 1$. Let us take example of length $4$. $4$ can be partitioned into $(0,4)$ or $(2,2)$ or $(4,0)$. There will be only one structure possible for $(0,4)$ as all of them are unpaired. For $(2,2)$, there will be $6$ different structure possible which can also be obtained by the calculation ${4\choose2} \times 1$. Similarly, for $(0,4)$ case, there are $3$ different possible arrangements which can also be computed by ${4\choose0}\times{3 \times 1}$. We have shown all $9$ possible permutation in Figure~\ref{fig:cases}. We are discarding the trivial case in which all the base are unpaired. For any $n$, this case will result in only one structure. That can be included by simply adding $1$ to the above formula.  Now, generalizing above results for any integer $n$ give us the following formula, 
$$No.\;of\;structures\;(\;\theta \in \mathbb{N})\;and\; n \ge 2  = $$
{\small
\begin{eqnarray}
\nonumber
&& \sum_{k=0}^{\theta - 1} \binom{n}{2k} \times \bigg\{ \big[ n - (2k+1)\big] \big[ n - (2k+1) - 2\big] \hdots 1  \bigg\} \\
\nonumber
&& \;\;\;\;\;\;\;\;\;\;\;\;\;\;\;\;\;\;\;\;\;\;\;\;\;\;\;\;\;\;\;\;\;\;\;\;\;\;\;\;\;n \in \mathbb{N}, \; n \; is\; even\; and\; n = 2\theta \\
\nonumber
&& \sum_{k=0}^{\theta - 1} \binom{n}{2k+1} \times \bigg\{ \big[ n - (2k+2)\big] \big[ n - (2k+2) - 2\big] \hdots 1  \bigg\}\\
\nonumber
&& \;\;\;\;\;\;\;\;\;\;\;\;\;\;\;\;\;\;\;\;\;\;\;\;\;\;\;\;\;\;\;\;\;\;\;\;\;\;\;\;\;n \in \mathbb{N},\;n\; is\; odd\; and\; n = 2\theta+1
\nonumber
\end{eqnarray}
}
Above formula can be simplified as follows:
\begin{theorem}
Let $S(n)$ be the number of secondary structures represented as a permutation of $n$ length. Then for $n \ge 2$, $S(n)$ is given by, 
$$ \sum_{k=0}^{\theta - 1} \binom{n}{m-1} \times \left\{ \prod_{p=0}^{(\frac{n-m-1}{2})} (n - m - 2p) \right\}  $$ Where if $n$ is odd then $n = 2\theta+1, m = 2k+2$. If n is even then $n = 2\theta, m = 2k+1$
\end{theorem}

In the Table~\ref{tab:permutations}, we have listed the results of above formula from $n = 2$ to $n = 9$. This series is often known as number of degree - n permutation of order exactly 2. There are alternative ways describe this formula. You can look at the values of this series upto $n = 300$ from web url \url{http://oeis.org/A001189/b001189.txt}.

\begin{table}[h]
\caption{No. of different secondary structure permutation possible different values of n}
\label{tab:permutations}
\begin{tabular}{c||c|c|c|c|c|c|c|c}
$n$&$2$&$3$&$4$&$5$&$6$&$7$&$8$&$9$\\
\hline
\hline
&&&&&&&&\\
$structures$&$1$&$3$&$9$&$25$&$75$&$231$&$763$&$2619$\\
\end{tabular}
\end{table}
%%%%%%%%%%%%%%%%%%%%%%%%%%%
\section{Classification of RNA}
Classification of RNA is a very subjective topic and it is very closely related with the database development of data. Many approaches have been taken to carry out this work. Some of the approaches (~\cite{citeulike:3124365},~\cite{citeulike:8436659}) are based on the biological aspect of the secondary structure whereas some (~\cite{citeulike:6365817},~\cite{citeulike:11428095}) are based on mathematics. Our permutation based approach also fall under the later category. In previous section, we talked about new hybrid notation, "permutSeq" which will contain eight different sequences as shown in Table~\ref{tab:newNotation}. Now, we will explain two simple algorithm we have developed to compute two types of similarity score and create clusters based on these two similarity score.
%%%%%%%%%%%%%%%%%%%
\subsection{Algorithm}
Algorithm~\ref{algo:similarityAlgo} computes two similarity score, first based on character sequence (containing A,C,G\&U) and the other based on permutSeq. "getSequence($i$)" procedure call on permutSeq returns $i^{th}$ character of an original sequence (row 1), "getIndicatorX($i$)" procedure call on permutSeq returns $i^{th}$ character of indicator sequence (row 5-8) of X (X = A, C, G \& U) and "getPermutation($i$)" procedure call on permutSeq returns $i^{th}$ character of permutation sequence (row 4).

\begin{algorithm}
\caption{Similarity Score Computation}
\algsetup{indent = 1em}
\begin{algorithmic}[1]
\STATE {\textbf{Procedure }(permutSeq P1, permutSeq P2)}
\STATE $Score1 \leftarrow 0 \; \backslash \backslash Strand Similarity Score$
\STATE $Score2 \leftarrow 0 \; \backslash \backslash Permutation Similarity Score$
\WHILE {Character i in P1 \AND Character j in P2}
\IF {P1.getSequence(i) = P2.getSequence(j)}
\STATE $Score1 \leftarrow Score1 + 1$
\ENDIF
\IF{P1.getPermutation(i) = P2.getPermutation(j)} 
\IF{P1.getIndicatorX(i) = P2.getIndicatoX(j) }
\STATE $ \backslash \backslash X = A,C, G \text{ and } U $
\STATE $Score2 \leftarrow Score2 + 2$
\ENDIF
\STATE $Score2 \leftarrow Score2 + 1$
\ENDIF
\ENDWHILE
\RETURN $ Score1, Score2 \backslash \backslash \; Score1, Score2 \in \mathbb{N} $
\end{algorithmic}
\label{algo:similarityAlgo}
\end{algorithm}

Algorithm~\ref{algo:clusterAlgo} describes procedure for creating clusters based on the similarity score data given as an input. Cluster creation can be applied only on a set of Similarity Score $S$ where,
\begin{eqnarray}
\nonumber
S&=&\left\{ (m,n) \mid m,n \in \mathbb{N}, \right. \\
\nonumber
&& \left. (m,n) = similarityScore( P1, P2 )  \right\}
\end{eqnarray}

Here, $SimilarityScore(P1, P2)$ procedure refers to output of Algorithm~\ref{algo:similarityAlgo} for inputs P1 and P2 whose type is permutSeq. We have given certain type of RNA to Algorithm~\ref{algo:similarityAlgo} and generated one single output file listing all permutSeq. If there are $n$ files in an input folder for Algorithm~\ref{algo:similarityAlgo}, it generates $^n\mathrm{C}_2$ number of permutSeq entries in output file which is used as an input for Algorithm~\ref{algo:clusterAlgo}.

\begin{algorithm}
\caption{Cluster Creation}
\algsetup{indent = 1em}
\begin{algorithmic}[1]
\STATE {\textbf{Procedure }(Score1[ ], Score2[ ])}
\STATE $clusters[\;]  \; \backslash \backslash \text{Array of Clusters} $
\FORALL {Integer i in Score1[ ] and Score2[ ]}
\STATE {$ \text{ Integer } j \leftarrow i $}
\FOR { j in Score1[ ] \AND Score2[ ]}
\IF {Score1[j] = Score1[i] \AND Score2[j] = Score2[i]}
\STATE $\text{Create Cluster Rule\_Score1[j]\_Score2[j]}$
\ENDIF
\STATE $ j \leftarrow j + 1 $
\ENDFOR
\ENDFOR
\RETURN $clusters$
\end{algorithmic}
\label{algo:clusterAlgo}
\end{algorithm}
Main features of the above two algorithms are:
\begin{itemize}
\item It can rank all types of sub structures including pseudonots.
\item Score computation runs in $O(n)$ which is much faster than any existing algorithm for doing similar kind of work.
\item It scores binding and binding with same base pairs in $1:2$ ratio.
\item It captures structural similarity between any two RNA secondary structure. 
\item It can be used to classify entire secondary structure data available or sub-classify existing classes of secondary structure.
\item It can be used to study the evolution of a particular species like HIV. For example, one can study which particular structural portions are not evolving among all different variants of HIV virus and infer which structural portions are actually tweaked by nature.
\item Right now we have done exhaustive analysis of a given data set but it can also be used to analyze entire data set with respect one particular strand.
\item Problem with the above point is that it is hard to analyze evolution if bases are being added or removed. It can be done only if modification of base is taking place.
\item One of the limitation is that it can be applied only on secondary structure data which is very less compared to actual sequence data. 
\item Similarity Score obtained are overall score, as of now there is no way to determine whether that score is result of contiguous portion of secondary structure or not.
\end{itemize}

\begin{figure*}[ht!]
\includegraphics{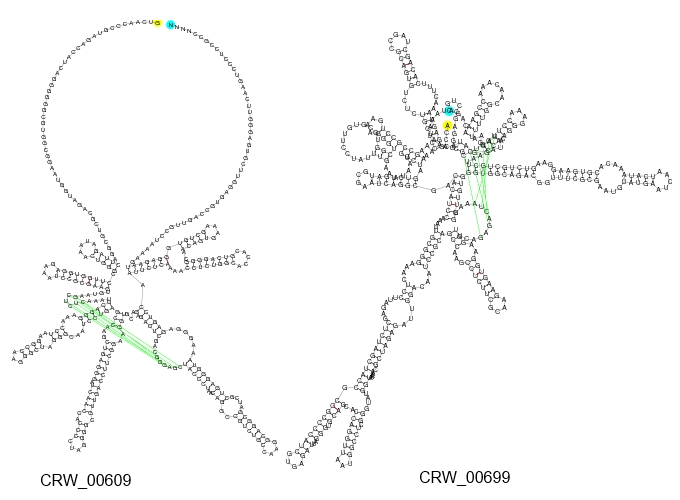}
\caption{Secondary Structures of CRW\_00609 \& CRW\_00699. They belong to a cluster having permutation and strand similarity score of 352 and 98 respectively.}
\label{fig:comparison}
\end{figure*}
%%%%%%%%%%%%%%%%%%
\subsection{Results}
We, have applied both of the above algorithm on various data available on~\cite{citeulike:3124365}. We have got some really interesting results which we have tabulated in Table~\ref{tab:results}. One very important point here is that set of all clusters created for any particular set of data is not a disjoint set because of the nature of the computation which is happening between every two entries from given data set. In Figure~\ref{fig:comparison}, we have shown two RNA secondary structure (accession number CRW\_00609 and CRW\_00699~\cite{citeulike:3124365}) side by side. From the given drawing of the structure it is very hard to find any similarity between these two structures. But both of them belong to the same cluster having strand similarity score of 98. Their total length are 373 and 429 and both of them contain a pseudoknot. Pseudoknot bindings are shown in green color in both the structure. Cluster in which they belong has a permutation similarity score of 352 which is very high score compared to total length of these two structres. 

Here let us clarify one thing, scores associated with any cluster are fixed and there can be many structures belonging to the same cluster but that does not mean all the members of a single cluster has the same similarity score with every other member of that cluster. But there are at least two members in the same cluster which can give exactly same similarity score as defined for that particular cluster. Another very useful observation regarding scores is that if permutation similarity score between any two structure is more than twice the strand similarity score than it is more probable that both the structures have similar sub structures.

\begin{table}[h]
\scriptsize
\caption{Clusters generated for different types of RNA}
\label{tab:results}
\begin{tabular}{p{1.6cm}||c|c|c|c}
\textbf{RNA Type}&\textbf{\# RNA}&\textbf{RNA Length}&\textbf{\# Clusters}&\textbf{Cluster Range}\\
\hline
&&&&\\
\raggedright Transfer Messenger RNA & 726 & 102 to 1331 & 13596 & 842\_421 to 38\_22\\
\raggedright Synthetic RNA & 450 & 4 to 302 & 590 & 172\_86 to 2\_1\\
\raggedright Signal Recognition Particle RNA & 394 & 12 to 533 & 4184 & 600\_298 to 2\_2\\
\raggedright 23S Ribosomal RNA & 205 & 18 to 4381 & 3536 & 5872\_2958 to 2\_2\\
\raggedright 5S Ribosomal RNA & 161 & 24 to 135 & 923 & 244\_121 to 2\_5\\
\raggedright Group I Intron & 152 & 14 to 2630 & 2236 & 1528\_263 to 2\_1\\
\raggedright Hammerhead Ribozyme & 146 & 40 to 119 & 451 & 236\_116 to 2\_5\\
\raggedright Other Ribosomal RNA & 64 & 12 to 500 & 196 & 232\_108 to 2\_2\\
\raggedright Other Ribozyme & 53 & 17 to 968 & 147 & 284\_142 to 2\_2\\
\raggedright Group II Intron & 42 & 27 to 2729 & 99 & 5040\_2520 to 4\_8\\
\raggedright Cis-regulatory element & 41 & 65 to 102 & 124 & 202\_100 to 24\_16\\
\end{tabular}
\end{table}
%%%%%%%%%%%%%%%%%%%%
\subsection{Comparison}
Now, we would like to compare our classification methodology which uses clusters and permutation with one of the very crude but yet useful type of classification~\cite{citeulike:6365817}. This is very famously known as RAG or RNA-As-Graphs. We are calling these classification very crude because it is loosing so much information about actual sequence like base count, type of base \& base pairing. It is useful because it is capturing all kind of structural information including pseudoknots along with their orientation in two dimensional geometry. Another reason for selecting~\cite{citeulike:6365817} is that this approach is using graph theory which is not a biological concept but a mathematical concept like ours. 

Their database contain majorly two types of graph called Tree Graph and Dual Graph. RNA structures having pseudoknots can not be represented by the simple tree graph which is why they have two major classification as tree graph and dual graph. In tree graph, they have classified motifs into nine sub category which is based on the number of vertex in the tree graph. Among these nine, smallest class is the one which has two vertex and there is only one possible topology for that configuration. On the other end, they have listed sub class having tree graphs with $10$ vertex. There are $106$ different possible graphs having $10$ vertex out of which only $10$ have been found so far. Whereas for dual graph they have sub category ranging from $2$ vertex to $9$ vertex.  Here, there are $38595$ different possible motifs for $9$ vertex, out of which only $4$ have been discovered till today. Now, if we compare this with our permutation based motifs, there are $2619$ different structures having length $9$. Currently, we do not have data or analysis to show how many of them have already been discovered. In cluster based classification, we have found $13596$ different clusters for the "transfer messenger RNA" type of RNA STRAND~\cite{citeulike:3124365} which is like a sub classification of biologically classified type of RNA which is not possible to do in RAG.
%%%%%%%%%%%%%%%%%%%%%%%%
\section{Future Work}
There are lot of possibilities one can think of related to the work that we have done. We will list down some of the idea that can be worked upon to extend our current work. Since, we are able to get unique representation in the form of permutation, one can explore many prospects of permutation. For example, one can define an operation on a permutation which will transform permutation into newer permutation and that newer permutation will uniquely represent a secondary structure if indicator sequences are known. One can experiment and obtain a heuristic results for selection of indicator sequence and its value to feed in after doing transformation on permutation. After all, two RNA secondary structure of related species are nothing but the transformation of each other which is driven by some unknown nature's law. This was mathematics based idea but one can also work on the software front of the current work.

A software based tool can be developed which can take any unknown (one dimensional sequence is known) RNA sequence (Let us call it $S$) and predict its secondary structure based on a cluster information. First, we can find similarity score of given sequence with entire database. Suppose there are $n$ entries in database of RNA secondary structure. We will find $n$  different similarity score based on this analysis. We choose the highest similarity score. Suppose sequence $S'$ is the one which was used for comparison to obtain highest strand similarity score. Now, we will search for the clusters which contain the same strand similarity score. In each matching cluster, we will search for the sequence $S'$, if found we qualify that cluster otherwise discard it. Finally, we will select the one or more structures from qualified clusters which are nearest in the size with given unknown sequence $S$. Currently, this is just a concept, we need to verify output of these procedure with the output of the prediction software like mFold~\cite{citeulike:100029} for the same unknown sequence input. 

Another software can be developed which takes any two sequences with known secondary structure and find six open reading frame portions. Cut all the ORFs and create a data set which can be used to do the exact same analysis that we have done for data sets of ~\cite{citeulike:3124365}. Similar work can also be done for two unknown sequence. Currently, our JAVA program implementation of two proposed algorithm does not show portions or index values of matching portions. A proper web based database can also be implemented which will classify all the data based on our permutation based approach and clusters will be put online for free public access. It can even be integrated for any existing databases for sub classification or it can be simply added to~\cite{citeulike:3124365} which has accumulated all different types of secondary structure databases.
%%%%%%%%%%%%%%%%%%%%
\section{Conclusion}
We have got some very interesting results in the form of clusters. For example, there are total $152$ Group I Intron secondary structure known and they are so tightly coupled that they have generated $2236$ clusters. Group II Intron class of RNA has total count of just $42$ known motifs but their size range from $27$ to $2700$ and its permutation similarity score varies from $4$ to $5000$. We have also been able to derive a formula that can count all possible secondary structure in terms of permutation. That can also be used to do the detailed analysis along with cluster data available for different sub classes of RNA.

%%%%% put algo
%%%%%%%%%%%%%%%%%%%%%%%%%%%%%%%%%%%
%\section*{Acknowledgment}
%The authors would like to thank Krishna Gopal Benerjee for useful discussions and Nikhil Agrawal for writing parts of the program for FR code enumeration and drawing some of the figures.
% trigger a \newpage just before the given reference
% number - used to balance the columns on the last page
% adjust value as needed - may need to be readjusted if
% the document is modified later
%\IEEEtriggeratref{8}
% The "triggered" command can be changed if desired:
%\IEEEtriggercmd{\enlargethispage{-5in}}

% references section

\bibliographystyle{IEEEtran}
\bibliography{ref}

% that's all folks
\end{document}